# A Multi-Agent Framework for Testing Distributed Systems

Hany F. EL Yamany, Miriam A. M. Capretz and Luiz F. Capretz
*Department of Electrical and Computer Engineering*
*Faculty of Engineering*
*University of Western Ontario*
*London, Ontario, N6A 5B9, Canada*
*{helyaman, mcapretz, lcapretz}@uwo.ca*

**Abstract**

*Software testing is a very expensive and time consuming process. It can account for up to 50% of the total cost of the software development. Distributed systems make software testing a daunting task. The research described in this paper investigates a novel multi-agent framework for testing 3-tier distributed systems. This paper describes the framework architecture as well as the communication mechanism among agents in the architecture. Web-based application is examined as a case study to validate the proposed framework. The framework is considered as a step forward to automate testing for distributed systems in order to enhance their reliability within an acceptable range of cost and time.*

## 1. Introduction

Today, large software systems are mostly distributed systems that are run on a loosely integrated group of networked processors [1]. Distributed systems are by nature more complex than centralized systems. This makes it more difficult to understand their emergent components and consequently their behaviour. As a result, testing and verification of these systems are complex. Furthermore, testing distributed systems is more challenging due to issues such as concurrency, fault tolerance, security and interoperability [1-3]. Therefore, testing distributed systems becomes a special problem which needs extraordinary solutions to cope with the distributed systems properties.

A number of authors [2, 4] have carried out research to test distributed systems that involve issues such as concurrency, controllability, security and timing. Other authors [3, 5] have focused their research on improving the performance of the testing process itself in order to make it faster and adaptive. Software reliability is the probability of execution without failure for some specified interval of natural units or time [6]. In order to obtain a more reliable distributed system, all of those works are needed to be integrated in a model that can work to enhance the functioning of the distributed systems and their testing process at the same time. Moreover, nowadays software testing is aiming to be more intelligent and self-managing and this requires automating it to decrease the cost and time of testing distributed systems.

To meet all the requirements outlined above, this paper introduces a novel multi-agent framework to test distributed systems. Generally speaking, an agent is an active computational entity that has relatively complete functionality and cooperates with others to achieve its designed objectives [7]. Agents are computer systems that are capable of independent, autonomous action in order to satisfy their objectives. As agents have control over their own behaviour, they must cooperate and negotiate with each other to achieve their goals [8]. The convergence of these agents' properties and distributed systems behaviour makes the multi-agent architecture an appropriate mechanism to improve the performance of distributed systems.

Agents in the proposed multi-agent architecture consist typically of two generic types: social (immobile) agents and mobile agents. Social agents are used to monitor the three-tier architecture of these distributed systems (i.e. server, middleware and clients) and to execute various scheduled testing types such as unit testing and integration testing. Moreover, mobile agents are used to carry out an urgent testing such as regression testing specified by a tester (i.e. human or an agent). This paper depicts these agents' functionality and describes their communication mechanisms. Since web-based applications have become a crucial component in the global information infrastructure and also, they are complex, hypermedia and autonomous distributed systems; they make good case studies to validate our framework.

The remainder of the paper is structured as follows. Section 2 shows the related work in testing distributed systems. Section 3 proposes the overall description of the multi-agent framework. Section 4 presents the agents' functionality and defines the communication mechanism between them. Section 5 explains the execution of the testing processes using the proposed collaborative multi-agent framework. Section 6 discusses a case study for testing web-based applications as a three-tier distributed system using the proposed framework. Section 7 includes the framework's evaluation. Section 8 provides conclusions and future work.

## 2. Related Work

This paper illustrates a framework for testing three-tier distributed systems. This includes the testing of a data repository in the server, middleware software and the different components on the clients' side. The proposed framework depends partially on the traditional testing techniques such as unit testing, integration testing, regression testing and stress testing. A combination of these testing approaches is applied to test the distributed system's reliability.

The literature on techniques for unit testing or other traditional testing is massive. In particular, unit testing has two major types [9]: 1) Control Flow Coverage Criteria and 2) Dataflow Coverage Criteria. Each one performs an important task in our suggested framework. Some papers [10, 11] focus on unit testing in their frameworks to test the single components only of distributed systems. On the other hand, Object-Oriented (OO) testing frameworks [2, 12] that include diagrams such as the sequence diagram and the class diagram are developed to test a graph of integrated components.

Two common generic strategies for testing concurrent programs are static and dynamic analyses. Static analysis techniques use a model of the program to be analyzed such as model checking, whereas dynamic analysis techniques collect information about the program through its execution [13]. For example, the work in [2] depicts a scenario including a combination of these strategies to implement the testing process. However, our approach uses a hybrid technique of these two strategies. A static analysis is used to generate an initial test suite that will be updated later using the dynamic analysis.

One common characteristic of all the existing testing techniques is that they are implemented sequentially. As a forward step in the testing mechanism, Lastovetsky [5] presents a new strategy to execute the testing process in parallel. This will accelerate the testing of complex distributed systems. This technique consists of the following steps:
1. Automatically partitions the input test suite into as many parallel streams as there are physical processors available (e.g. clients in three-tier system).
2. Launch all the streams in parallel and wait until all these streams of test cases are complete.
3. Create a final report on the parallel execution including the feedback resulting from the implementation of these streams.

The papers [7, 14] use multi-agent systems as a dynamic tool that can improve the testing of distributed systems such as web services. Their major idea of employing multi-agent systems is to continuously monitor the changes that might occur in the structure of web services and dynamically produce the suitable testing technique accordingly. Also, agents can verify the testing results. This mainly helps in enhancing the performance of these systems. In addition to that, our proposed framework monitors the user usage in order to increase the leverage of the testing process by increasing the chances to discover most of the defects that might appear in both the server and clients sides.

Not many papers on distributed systems testing deal with the issue of reliability. For example, Musa [15] creates an operational profile that can be used as a basis to reveal system defects while Levendel [16] uses a mathematical model for defect removal. However, these two techniques are more effective in testing communication systems only.

## 3. System Description

Three-tier distributed system architecture consists of a server, middleware and multiple clients. The server contains the data repository of the distributed application, whereas the middleware is considered to be the software bus associated with those clients. Figure 1 depicts this structure.

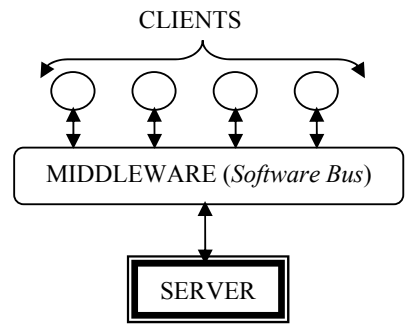

**Figure 1: The Architecture of a 3-tier Distributed System**

The proposed multi-agent framework copes with the distributed system structure. The framework consists of three levels of autonomous and adaptive agents. Figure 2 illustrates the framework. The first level of agents is on the server side. Basically, it is a single agent that monitors the data of the distributed application and is called the Database Repository Agent (DRA). The second one – Middleware Controller Agent (MCA) – is located at the middleware and is the kernel of the proposed framework. Its main goals are to investigate the middleware behavior, collect the return feedback from the clients and make an integrated report about the system. Finally, a group of social agents is distributed over the available clients. Each one is named Client Checker Agent (CCA). This group of agents is coupled with the distributed application components and work in parallel with those components to examine their functionality.

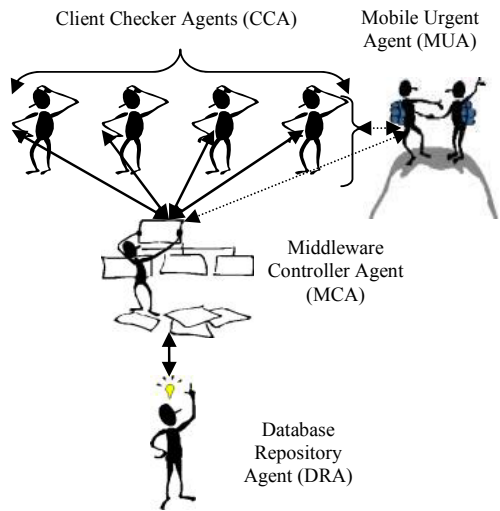

**Figure 2: Multi-Agent Framework For Testing Distributed Systems**

The framework can be extended to execute more testing procedures at the request of the tester. In some crucial unexpected behavior of a distributed system, the tester can ask for further testing and this can be done by sending a supportive mobile agent that could help in that mission. This agent's name is Mobile Urgent Agent (MUA). Details of the proposed framework are discussed in subsequent sections.

## 4. The Multi-Agent Architecture

An agent is a piece of software that can be viewed as perceiving its environment through sensors and acting upon that environment through effectors [17]. Agents are autonomous, intelligent, flexible, cooperative and reactive [18]. These properties can be described as follows:
1. *Autonomy:* Agents automatically monitor a distributed system during its runtime.
2. *Intelligence:* Agents reveal most of defects that might occur by performing the different testing techniques. As a result, they generate fresh test cases that enhance the performance of testing execution.
3. *Flexibility:* Agents perform different testing approaches depending on the changes and the development of distributed systems.
4. *Cooperation:* Agents communicate with each other in the proposed framework and consult with the testers to ensure the valid execution of the different testing techniques.
5. *Reactivity:* Agents reclaim any found error when misbehavior of a distributed system occurs.

**TABLE 1: Agents Functions**

| Agent Name | Goal | Perception | Action | Output |
|---|---|---|---|---|
| DRA | Monitor the distributed system's data and maintain the database of test suites. | Initial test suites from the tester. Modified test suites from MCA. The design document of the distributed system. | Check the application's data organization and update various test suites. Generate expected output results for the testing. | A report for the tester. Test suites for future testing processes. |
| MCA | Unit Testing for Middleware. Integration testing with the clients for the whole system. System testing. | Test suites from DRA. CCA's feedbacks. Instruction from the tester. New test suites from CCA. | Execute unit and integration testing. Release mobile agents to help CCA. | A general report about all completed testing processes. |
| CCA | Unit testing | Instruction and test suites from MCA. Users' log files. | Implement unit testing. Trace users' log file. Discover defects and new test cases if available. | A report to MCA. New test cases if available. |
| MUA | Urgent testing such as stress testing. | Instruction and test suites from MCA. Users' log files. | Execute the required test according the tester's needs. | A report to MCA. New test cases if available. |

Software agents as shown above can be a suitable solution to automatically perform the testing of distributed systems.

### 4.1 The Agents' Functionality

To exhibit the agents' functionality in this framework, their goals, perceptions, actions and outputs are demonstrated according to the former agent definition. Table 1 describes these activities.

At the first level of this framework, the Database Repository Agent (DRA) is responsible for checking the data of the distributed application. Also, DRA builds a database to store the different data suites that are used in the various testing techniques. Concurrently, it generates the expected output that can be derived from the design document of the distributed system to evaluate the results of these testing processes. It periodically filters useless test cases in its database in case of system maintenance caused by adding new functionalities to the system. The duties of the Middleware Controller Agent (MCA) are: to execute the unit testing for the middleware software; to perform the integration testing with the clients and the server; and finally, to release mobile agents to perform more different testing approaches in case new testing is needed or for another testing approach that is not defined in the Client Checker Agent's schedule (for example doing a regression test after updating the distributed system). Finally, the Client Checker Agent (CCA) implements the unit testing on the clients' side. In non-testing periods, it monitors the log file of a user's usage of the distributed system, reveals any defects and generates new test cases as needed.

## 4.2 The Communication Mechanism between the Agents

Due to the diversity of the environments where the distributed systems are running, the agents use SOAP messages to communicate with each other. The data in these SOAP messages vary depending on the testing technique's type. For example, MCA sends SOAP messages to the CCAs to perform unit testing with specified test cases; these SOAP messages include the testing type (eg. unit testing) and the test cases which CCAs use in the required testing.

Another example includes when a CCA finds a new test suite, it sends a SOAP message to MCA claiming that a new test case is found. This SOAP message includes the discovered defect, the test case that causes this defect and the operation name in which the defect occurred. MCA then forwards this message to DRA to check the test case. DRA compares the test case with the test cases in its database. DRA uses the defect type and the operation name as metric in the comparison process. Subsequently, it discards the new test case if it is similar to an existing one. Otherwise, it adds it to the database.

## 5. Agents for Testing Distributed Systems

Testing a distributed system is a costly and time consuming process. Therefore, the agents in the proposed framework are intelligent; they monitor the behavior of the distributed system and they execute any required testing in case a defect is found or misbehavior is checked. Furthermore, all agents in this framework are autonomous and work in parallel which constitute an essential factor in order to decrease the time and cost. The various testing techniques can be carried out as follows.

## 5.1 Unit Testing

Both MCA and CCA execute the unit testing: MCA tests the middleware and CCA tests the components of the distributed system at each client separately. Because the kernel of the distributed software resides in the middleware and so MCA uses Dataflow Coverage Criteria [9] to analyze the structure and then the behavior of the middleware software. However, CCA use Control Flow Coverage Criteria [9] (i.e. White-Boxing Testing) and this is because the running components inside the different clients are smaller in size and testing time is therefore reduced.

## 5.2 Integration Testing

After MCA and CCA finish the execution of the unit testing, MCA sends SOAP messages to CCAs and also to DRA to begin the implementation of integration testing to examine the whole system architecture.

## 5.3 Regression Testing

The system tester may ask MCA to perform regression testing after the system is updated or when new operations are added to the system. As in integration testing, MCA sends SOAP messages to all agents in the framework to carry out this kind of testing. If the agents are busy executing other testing processes, MCA releases MUA to perform this testing instead of the busy agents.

## 5.4 Stress Testing

One major goal of the proposed software testing framework is to enhance software reliability. System testing is an example of one of the techniques used to enhance software reliability. System testing has many types; one of which is stress testing. The agents in our proposed framework can perform stress testing to measure the number of defects per estimated period of time. The testing is done as follows: MCA asks DRA for suitable test cases to perform stress testing. DRA sends SOAP messages that include the required test cases that may be automatically generated. Consequently, MCA performs stress testing to count the defects in order to measure the reliability. Finally, MCA produces a detailed report for the tester that includes the results of the testing.

## 6. Case Study: Testing Web-Based Applications

A three-tier web application consists of three separate layers [19] which are shown in Figure 3.

1. A front end which is composed of many different clients.
2. A middle dynamic content processing which is the middleware. For example, Java EE platform.
3. A backend Database comprising the data sets and the Database management system software that manages and provides the access to the data.

The reliability of the web-based applications can be tested by examining each layer separately and then checking the whole architecture as an integrated unit by testing the interaction between these layers. Pressman [20] states some factors that should be considered during testing the web's reliability such as correct link processing, errors recovery and user input validation.

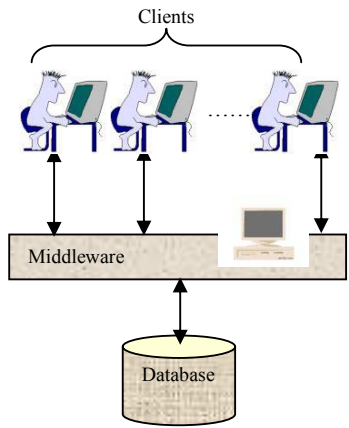

**Figure 3: Three-tier Web Applications Architecture**

The proposed framework for testing these web-based applications in three different cases is assessed here. They include a link failure, a user input validation and an error in retrieving user data.

### 6.1 Case 1: Link Failure

When a user presses a hyperlink in a page on the client side which generates an error (e.g. it does not open the linked page), the CCA records this error and then sends it in a SOAP message to MCA to reclaim the error. The SOAP message includes the page title as the operation name and the hyperlink name as the defect type as mentioned before in the communication mechanisms. Consequently, MCA generates a report for the tester that points out this defect. MCA sends another SOAP message to DRA to try to produce a new test case if possible.

### 6.2 Case 2: User Input Validation

When a user input some data (e.g. in E-commerce websites), an error might occur during data validation. In this case, CCA also registers this error and sends it in a SOAP message to MCA. The SOAP message contains the submit operation as the operation name, registration defect as defect type, from name and the page name that includes this form. This additional information helps to estimate precisely the defect that has occurred. As usual, MCA produces a report about this defect and sends this case to MRA to generate test cases that can be used in the future when executing the unit testing.

### 6.3 Case 3: A Problem in Retrieving User Data

Sometimes, a user attempts to retrieve data and a problem arises. In this case, MCA and DRA work together and exchange SOAP messages to analyze this problem and to determine whether its causes lie in the retrieval function in middleware software or in inconsistent data in the database or other reasons (e.g. network failure that is beyond of this paper scope). In similar cases, MCA asks to perform integration testing to reveal this error and it may release MUA to help. Finally, DRA produces an expected result for the tested function (i.e. the retrieving function) from the design document to compare it with the testing result which in turn validates the testing process.

## 7. Framework Evaluation

The proposed framework is split into three separate levels. All of them execute the testing process in parallel. In addition, the agents on the client side also work in parallel which decreases the testing time. The framework is not complex which helps to minimize the testing costs.

The agents in this framework automatically monitor the system and they are naturally autonomous. This can increase the level of detection of any defects that may occur. This can be achieved as mentioned above by reading the users' log file and recording any misbehavior from the system.

The proposed framework can execute different testing approaches depending on the distributed system behavior. It can also perform unexpected testing techniques that are not defined in their schedule by releasing MUA which can dynamically perform the new testing.

Finally, the agents can generate fresh test cases during the system monitoring in both sever and client sites and update its database in order to improve the testing process. This makes our framework more beneficial than the suggested frameworks in [7, 14]. It can also produce expected results for the distributed system functionality to compare with the results from the testing process which helps in the evaluation of the various testing techniques. Basically, this evaluation comprises excellent data that are examined by the tester and the agents to improve the 3-tier distributed system testing.

## 8. Conclusions and Future Work

In this paper, we have proposed a new multi-agent framework to test 3-tier distributed systems. This framework includes intelligent agents that work together in parallel in order to decrease testing cost and time. A framework is described and the communication mechanism between them is also defined.

Furthermore, this framework is able to perform different testing techniques according to the distributed system behavior. It can generate new test cases by monitoring the users' actions and it creates expected outputs from the execution of the various components in a distributed system to compare it with the testing results. This helps to validate the testing processes.

A case study for testing web-based applications as three-tier distributed systems using the proposed framework is discussed. A scenario that simulates the expected framework behavior to test websites in three different cases is studied. The scenario proves the efficiency of the proposed framework in recovering any error that occurs in each layer of the three-tier web applications architecture and even in the integration work between them.

This framework can be considered as a step toward automation of the distributed systems testing in order to enhance their reliability within an acceptable and reasonable range of cost and time. Since this study proposes the basic framework and related functionalities, in the future we will develop a tool to validate this model and make further testing in other distributed systems to optimize the model's functionality.